# Fully-control of OAM vortex beam and realization of retro and negative reflection at oblique incidence using dual-band 2-bit coding metasurface

Hamza Ahmad Madni[1,2,†], Shahid Iqbal[1,2,†], Shuo Liu[1,2], Lei Zhang[1,2], Tie Jun Cui[1,2,*]

[1]State Key Laboratory of Millimeter Waves, Southeast University, Nanjing 210096, China.

[2]Synergetic Innovation Center of Wireless Communication Technology, Southeast University, Nanjing 210096, China.

[†]These authors contributed equally to this work.

[*]Corresponding Author: tjcui@seu.edu.cn

**Abstract**

This paper addresses a reflection-type dual-band 2-bit coding metasurface (CM) design to achieve the dual-band functionalities in two different operating bands, independently. We are particularly interested to control linearly-polarized incident waves by encoding the propagation / dynamic phase in order to realize and fully control the orbital angular momentum (OAM) vortex beam (VB) for dual-band, independently. In this regard, we perform digital addition operations to combine the coding sequences of traditional OAM and phase-gradient to construct the proposed CM. The proposed CM generates and steers the OAM-VB with different topological charges ($l$) in both lower ($f_l$) and higher ($f_h$) bands, independently. Similarly, the proposed CM is further extended for OAM-VB shaping by combining the coding sequences of traditional OAM and splitting phase unit-cells, and, we discuss two different scenarios. Firstly, OAM-VB splitting is performed to achieve different '$l$' in both lower and higher bands, independently. Secondly, quad-beam shaping is realized to verify the proposed concept of this paper. Finally yet importantly, we also examine that the reflected beam directions can be further independently controlled to achieve anomalous, negative and retro-reflection at two different operating bands. The proposed concept will open up possibilities of multi-functional meta-devices and will be useful in many applications such as optical spanners, quantum optics, spatial mode multiplexing for telecommunications, and astronomy etc.

**1. Introduction**

Orbital Angular Momentum (OAM) carried by vortex beam (VB) is increasingly recognised as an important dimension in the field of communication optics. The phase profile of OAM is described by the '$\exp(il\phi)$', in which the topological charge '$l$' can be any integer value [1–4]. This unlimited freedom of '$l$' provokes the larger information capacity. Thus, OAM-VB has various potential applications in antennas and radio communication to increase the communication capacity without broadening the bandwidth [5, 6]. The generation of OAM-VB has been investigated in the microwave bands [7–10] and optical domain [11–13]. In addition, several good approaches exist for OAM-VB generation including circular phased array antenna [14–17], spiral parabolic antenna [18–20], and spiral phase plate (SPP) [21, 22]. Recently, metasurfaces (MS) provided a novel route to successfully generate the OAM beams [23–36] having more advantages compare to the conventional methods such as: easy-to-fabricate, light weight, and ultrathin profile [37, 38] etc.

In general, MS acts as a lens or a reflector in OAM-VB generation. In which the transmission and (or) reflection phase of the incident wave can be easily tailored to satisfy the phase-condition of OAM-VB by properly adjusting the rotation angle / size of the MS's elements. Curiously, traditional MS does not resolve the basic problem to generate flexible and multiple OAM-VB with different angles. To overcome this, the coding-metasurface (CM) [39] concept was proposed that creates link between the digital codes and MS particles to realize the powerful manipulation of electromagnetic (EM) waves using simple coding sequences. Since then, various functionalities [39–48] have been obtained by simply implementing the corresponding coding sequences, i.e., beam focusing [43, 44], beam steering [42–44], random EM-wave scattering [47], beam shaping [39,48] and OAM-VB generation [49–51] etc. Further ahead, passive MS [43,44,48,52,53] were investigated to fulfil the demand of multi-functional devices. On the other hand, these passive MS designs are hard to be electronically controlled. In order to address these flaws of passive MS, active MS [54–56] approach is employed involving multi-band and multi-functional applications. Similarly, more endeavours have been made to improve the bi-functional MS opportunities for EM waves controlling in full-space [57].

However, the previous works done dealt with the bi-functional realizations for steering and shaping of EM beams. Alongside this, Lei Zhang et al. proposed a single-band Pancharatnam-Berry (PB) / geometric-phase CM to steer and shape the OAM-VB by controlling the circularly polarized EM waves [51]. It is clear that linear polarization is an important factor to use OAM-VB in many applications for example: optical spanners [58], quantum optics [59], spatial mode multiplexing for telecommunications [60], and astronomy [61–65] etc. Therefore, multi-functional dual-band CM for linear polarization still needs to be explored for the generation of flexible, and multiple OAM-VB with different topological charges, independently.

So the work in this paper focuses on the realization of flexible and multiple OAM-VB with different modes at two different frequency bands. Compare to the PB-CM design [51], here, two different kinds of the phase modulation apertures are closely integrated into one meta-atom, and then use the whole CM aperture to realize the OAM beam independently at dual-band for linear polarization. We combine the normal OAM coding sequence and phase gradient coding sequence to design a new coding sequence of the proposed CM that can operate at two different operating bands independently, i.e., lower ($f_l$) and higher ($f_h$) bands. By independently designing dynamic / propagation phase distributions, the proposed CM is expected to achieve OAM-VB with different topological charges ($l$) and deflect to opposite half planes. This phenomenon of OAM-VB deflection can be seen in Fig. 1a. We further extend this work to generate OAM-VB shaping and discuss two different scenarios. In the first case, OAM-VB splitting is performed to achieve different topological charges ($l$) in both lower and higher bands, independently. Similarly, quad-beam shaping is realized to verify the proposed concept of this paper. Overall, this OAM-VB shaping task is accomplished by the addition of OAM-phase and splitting phase [39,48] unit-cells. Finally yet importantly, Fig. 1b illustrates that the reflected beam directions can be further independently controlled to achieve anomalous and retroreflection at two different operating bands. The proposed concept will be crucial in the design of multi-functional meta-devices with multi-spectral features in terahertz and optical regimes.

The rest of the manuscript is organized as follows: Section II presents relevant aspects of the unit-cell design that is important in metasurface formation. The detailed discussion of proposed designs and its various applications are examined in Section III. Section IV concludes the manuscript.

## 2. Materials and Methods

### 2.1 Unit-cell design

To start with, we required sixteen quantized reflection phases to achieve the dual-band 2-bit functionality of the proposed CM that obey the binary rule of $2^{(m,n)}$. Where, '$m$' and '$n$' represent the number of operating bands and the number of quantized phases, respectively. In the meanwhile, the quantized phases for each operating band can be found with the help of following matrix, given in Eq. (1).

$$M^{2-bit} = \begin{bmatrix} 00/00 & 00/01 & 00/10 & 00/11 \\ 01/00 & 01/01 & 01/10 & 01/11 \\ 10/00 & 10/01 & 10/10 & 10/11 \\ 11/00 & 11/01 & 11/10 & 11/11 \end{bmatrix} \qquad (1)$$

In Eq. (1), each binary number before slash (in each row) represents the phase state for lower band ($f_l$). On the other hand, the binary numbers after slash show the phase state in the higher band ($f_h$). The unit-cell of the proposed CM is composed of two square and (or) rectangular metallic patch resonators with a ground metallic layer for complete reflection, can be seen in Fig. 2a. Therefore, two dielectric spacers F4B [with $\varepsilon = 2.65$ and tangent loss $\delta = 0.001$] are used to separate the ground metallic layer and two patches. To make the designing procedure easy to understand, the geometric values of coding particles are given as following: lattice constant $p = 7mm$, thickness of each dielectric substrate $h = 1mm$, thickness of ground layer and each copper metallic patch $t = 0.018mm$. In addition, $w_1$ and $w_2$ describe the side lengths of both inner and top metallic patches along x-axis, respectively. Similarly, $l_1$ and $l_2$ represent the side lengths of both inner and top metallic patches along y-axis, respectively.

The specific aim of this paper is still questioned to provide a dual-band 2-bit CM in favour of controlling OAM-VB for linear polarization. In this context, we choose the off-diagonal phase states of matrix '$M^{2-bit}$', i.e., $M_1 = [00/11, 01/10, 10/01, 11/00]$. Interestingly, it can be noticed that the direction of phase gradient is opposite in the two operating bands. For example, the phase gradients for lower-band $[00, 01, 10, 11]$ are in ascending order, while for higher band, the phase gradients $[11, 10, 01, 00]$ are in descending order. Therefore, such difference in phase gradients provoke steering in the opposite half-space by sharing the same aperture and this phenomenal beauty is the key point to accomplish the proposed concept of this paper.

Here, we claim that the proposed dual-band 2-bit CM determines the properties of the OAM-VB generation and steering in opposite half-plane with equal and opposite topological charges, in the presence of linear-polarized incidence wave. In this scenario, all sixteen coding-particles are obtained from '$M_1$' by optimizing the dimensions $(l_1, w_1, l_2, w_2)$ of two patches as shown in Fig. 2a. Moreover, commercially available software CST, MWS is utilized to verify the validation of the proposed unit-cell under normal incidence of $x$-polarized plane waves with periodic boundary condition and Floquet port excitations. Optimized geometrical parameters for all coding particles of Fig. 2a are achieved by employing parametric sweeps for the values of the side-lengths of both patches in the unit-cell. The desired reflection-phases for the acquired coding particles are achieved in both the lower-band $(9.2-10 GHz)$ and higher-band $(14.4-15.1 GHz)$ under the normally incident $x$-polarized wave, can be seen in Fig. 2b and 2c, respectively. It can also be noticed that for each four identical reflection phases in lower-band, there exist four reflection phases in higher-band. Therefore, two consecutive phases are $90°\pm10°$ out of phases from each other to mimic dual-band 2-bit coding particle. Therefore, Fig. 2b illustrates the phase response of first four meta-atoms in which the pink coloured shaded region represents the lower-band and blue-colour is for higher-band. Similarly, phase responses of all coding particles in lower-band can be seen in Fig. 2c. Moreover, phase responses of all coding particles in higher-band are given in Fig. 2d. Most importantly, phase responses of both the diagonal and off-diagonal particles of matrix '$M^{2-bit}$' for both operating bands are presented in Fig. 2e and Fig. 2f, respectively. The amplitude responses of all coding particles in both operating bands are given in Fig. 2g.

## 3. Results and Discussions

Firstly, the concept of super unit-cell is used for various coding sequences to avoid unwanted coupling between different coding particles that are placed adjacently and have different patch sizes. Secondly, the super unit-cell has freedom for the coding sequence to achieve the desired functionalities in different operating bands, independently. As example, here we focus on the generation of OAM-VB in different direction, and to achieve negative, anomalous and retro-reflection, independently. In the following, full-wave simulations are carried out by employing CST, MWS with open add space boundary condition under both normal and obliquely incident $x$-polarized plane waves.

### 3.1. OAM-VB steering

Although, a series of research has been done for beam steering using MS [39,41–43,66–68] and such previous designed meta-devices are applicable for single-band. Meanwhile, OAM-VB steering has also been investigated in ref. 51 having lack of multi-band and linear-polarization features. Here, we demonstrate OAM-VB steering by designing dual-band 2-bit CM with the coding particles of $M_1$ in the proper coding sequences. In this perspective, the convolution theorem [69] helps to link the coding patterns of different functionalities that enable us to construct the coding pattern of new design of CM. In easy words, the new coding sequence of the proposed CM is the combination of both the normal OAM coding sequence and phase gradient coding sequence. For the current design, the coding sequence has the binary pattern of '$M_3 = M_1 + M_2$', where the addition is the binary addition, and '$M_2$' is the phase gradient coding sequence, i.e., $M_2 = 00,01,10,11$.

To verify the OAM-VB steering feature, the proposed concept of unit-cell (Section 2.1) is adopted. Overall, the full structure is composed of $32 \times 32$ coding particles, where the period of gradient coding sequence is 16 unit-cells under *x*-polarization incidence for both operating bands. Moreover, the deflection angle for the proposed design in two operating bands can be calculated by Eq. (2).

$$\theta_d = \sin^{-1}\left[\lambda/\Gamma\right] \qquad (2)$$

where $\lambda$ is the wavelength of the incident wave and $\Gamma$ is the period of the coding sequence.

For easy to understand, Fig. 3 depicts the coding patterns of proposed design for linearly *x*-polarized incidence. Fig. 3 contains three different coding patterns (Figs. 3a−c) and their corresponding 3D and 2D scattering patterns for both lower-band (Figs. 3d−e) and higher-band (Figs. 3f−g). The coding pattern in Fig. 3a consists of four segments and the successive phase step from one segment to another is $\pm 90°$. This kind of rotated phase distribution has tendency to generate OAM-VB of mode $l = 1$. In the next step, we add the coding pattern '$M_1$' of Fig. 3a with the phase gradient coding sequence '$M_2$' varying along the *x*-direction (Fig. 3b). The resultant coding pattern '$M_3$' can be seen in Fig. 3c that generates the OAM-VB with deflection angle $\theta_d$. For lower-band, the generated OAM-VB from Fig. 3c deflects to the left-side that can be seen in Figs. 3d−e. Similarly, for higher-band, Figs. 3f−g represent the generation and deflection of OAM-VB towards right side. It can be noticed that the topological charge of the realized OAM have opposite values in each operating bands, such as '$l = 1$' in lower-band and '$l = -1$' in higher-band.

*3.2. OAM-VB shaping*

From aforesaid discussion, the deflection of OAM-VB to any direction is achieved by addition of two different coding patterns. Similarly, beam shaping [39,48] of OAM-VB can be obtained by adding the beam-shaping coding sequence with the coding sequence of normal OAM-VB. In this section, we will again use the convolution operation to realize flexible and multiple OAM-VB for linear-polarization.

For this purpose, firstly we add the OAM-VB coding pattern (Fig. 3a / Fig. 4a) to the coding pattern of Fig. 4b, which has the tendency to split the reflected waves into two equal beams with the deflection angle of $\pm\theta_d$. The newly designed coding pattern can be seen in Fig. 4c, which has the ability to generate two OAM-VB from the same aperture deflected at equal and opposite angles. The simulated results for this scenario are shown in Figs. 4g,h,k,l. Figs. 4g and 4k represent the scattering pattern of 3D and 2D for lower-band with x-polarized incidence, respectively. On the other side, Figs. 4k and 4l represent the scattering pattern of 3D and 2D for higher-band with *x*-polarized incidence, respectively.

Moreover, the coding Pattern of OAM-VB (Fig. 3a / Fig. 4d) is added to the coding pattern of chess board configuration (quad-beam generator) as shown in Fig. 4e. Whereas, in the chess board configuration, the reflected field split into quad-beam. Consequently, the new coding pattern (Fig. 4f) is the addition of quad-beam generator and the OAM-VB that has the ability to generate quad-OAM-VB. The simulated results for this case can be observed in Figs. 4i,j,m−p. Fig. 4i shows the 3D scattering pattern of OAM-VB quad-splitting in the lower-band for *x*-polarized incident. In which the normalized incident wave creates four different OAM-VB in opposite half-planes. 2D scattering pattern of OAM-VB quad-beam in lower-band is given when the cut angle $\theta = -45°$ (Fig. 4m) and $\theta = 45°$ (Fig. 4n). Fig. 4j represents the 3D scattering pattern of OAM-VB quad-splitting in the higher-band for x-polarized incident. In addition, 2D scattering pattern of OAM-VB quad-splitting in higher-band is given when the cut angle $\theta = -45°$ (Fig. 4o) and $\theta = 45°$ (Fig. 4p). In all cases, as the direction of phase gradient of the OAM coding sequence in the two operating bands is opposite thus, the topological charge of the realized OAM in the two operating bands have opposite values, i.e., '$l = 1$' in lower band and '$l = -1$' in higher band. The simulation results have an excellent agreement with the theoretically predicted results.

### *3.3. Negative and retro-reflection*

All the positive results for the above-described proposed CM in Section 3.1−3.2 point out the fully-control of OAM-VB in different bands, simultaneously. In some scenarios, we need to control the reflected beams to achieve anomalous and retroreflection etc. For this sake, CM provides more flexibility to deflect the reflection or transmission beam in the user defined direction [41–43] for both normal and oblique incidences. In addition, the retroreflection phenomenon has also been investigated using planar MS [70] in which the angle of reflection is same as the angle of incidence. However, these designs are applicable for predefined functionality and still suffer multi-functional and multi-spectral properties. Similarly, a need still exists to achieve anomalous and retro-reflection with oblique incidence of linear polarization.

In general, retroreflection phenomenon can be realized using a bi-layer planar MS; luckily our design is also a bi-layer structure which can be used to realize this important functionality. Here, we are supposed to investigate a CM to achieve anomalous and retroreflection with oblique incidence at different operating bands independently, while the aperture remains unchanged. The proposed CM has the ability to control the co-polarized reflected waves to different angles based on the period and predesign selection of the coding sequences. In this regard, a CM is designed encoded with the off-diagonal elements of the matrix '$M^{2-bit}$' with a period of eight coding particles having the coding sequence, 00/11,00/11,01/10,01/10,10/01,10/01,11/00,11/00 00/11,00/11...... The overall size of the design is $32 \times 32$ particles with an area of $224 \times 224 mm^2$.

Fig. 5 demonstrates the simulation results for 3D and 2D far-field patterns of the proposed CM to realize anomalous, negative and retroreflection for *x*-polarized oblique incidence. The anomalous reflection in lower-band for a series of incidence angles ranging from 9° to 11° along with the retroreflection in higher-band is achieved. For brevity, here, we show only the results at an incidence of 10° only, whose 3D scattering patterns in the two operating bands can be seen in Figs. 5a and 5c, respectively. The corresponding 2D equivalents can be seen in Figs. 5b and 5d, where a retroreflection in the higher-band can be seen at 10° (Fig. 5d). Furthermore, the reflected beam direction can be controlled by tuning the incident angle of the *x*-polarized wave, and in this way negative reflection for a series of oblique incidence angles can be achieved. Here, we show the anomalous reflection realized for incidence angle 7° in lower-band along with the negative reflection in higher-band. The simulated results for this case can be seen in Figs. 5e and 5g, respectively. Moreover, Figs. 5f and 5h show the 2D radiation patterns in lower-band [for anomalous reflection] and higher-band [for negative reflection], respectively.

## 4. Conclusion

We presented a reflection-type dual-band 2-bit coding metasurface (CM). We combined the coding sequences of traditional orbital angular momentum (OAM) vortex beams (VB) and phase-gradient to construct the proposed CM. The proposed CM can generate and steers the OAM-VB with different topological charges (*l*) in both lower ($f_l$) and higher ($f_h$) bands, independently. Multiple OAM-VBs have also been realized by applying the addition operation on the coding sequences of beam-splitting (and quad-beam shaping) and normal OAM-VB. Moreover, reflected beam directions are independently controlled to achieve anomalous, negative and retroreflection at two different operating bands. The proposed concept will open possibilities of multi-functional meta-devices with multi-spectral properties for reflection type optical devices.

**Authors Contribution.** H. A. Madni and S. Iqbal designed the devices and carried out the simulations. S. Liu, L. Zhang, and T. J. Cui analyzed the data and interpreted the results. H. A. Madni drafted the manuscript with the input from the others. T. J. Cui supervised the project.

**Funding.** This work was supported in part by the National Key Research and Development Program of China (2017YFA0700201, 2017YFA0700202, 2017YFA0700203), in part by the National Natural Science Foundation of China (61631007, 61571117, 61501112, 61501117, 61522106, 61731010, 61735010, 61722106, 61701107, and 61701108), and in part by the 111 Project (111-2-05). H. A. Madni acknowledges the support of the Postdoctoral Science Foundation of China at Southeast University, Nanjing, China, under Postdoctoral number 201557.

**Acknowledgments.** The authors would like to thank the Editor and the anonymous reviewers for their insightful comments and constructive suggestions that certainly improved the quality of this paper. H. A. Madni thanks S. Iqbal and S. Liu for critical discussions. H. A. Madni would like to express his gratitude to Prof. Tie Jun Cui who welcomed H. A. Madni to his research group when H. A. Madni was at the crossroad of continuing his research career.

**Conflicts of Interest.** The authors declare no conflict of interest.

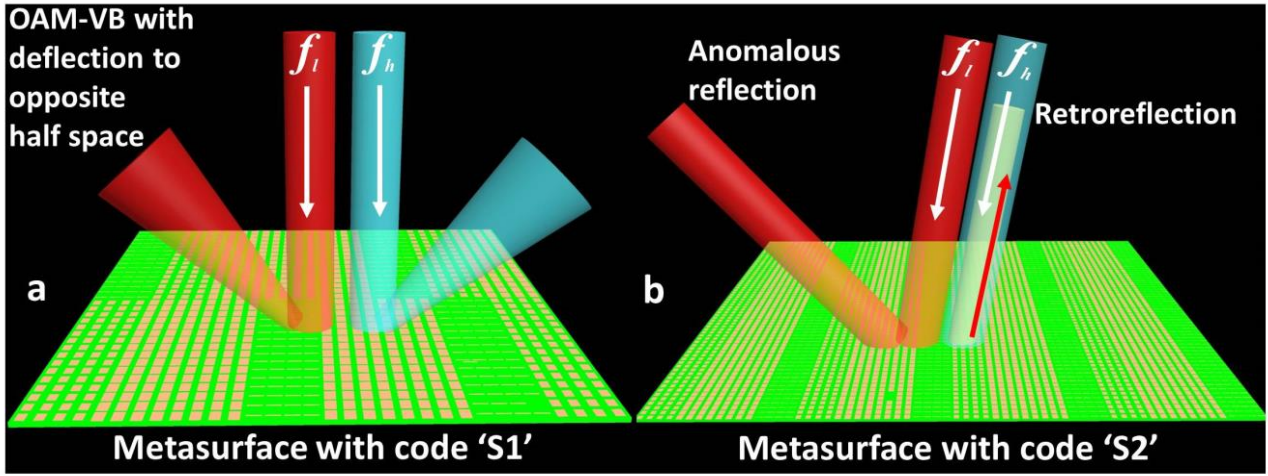

**Figure 1:** Schematic illustration of the dual-band 2-bit CM to generate and fully control both the: OAM vortex beam, and anomalous- and retro- reflection, with linear polarization and oblique incidences, respectively. **(a)** Independent manipulations of incident waves to generate and deflect OAM-VB to opposite half-plane in both lower ($f_l$) and higher bands ($f_h$). **(b)** Independent manipulations of planar waves to achieve anomalous reflection (in lower band) and retroreflection (in higher band) under the oblique incidence.

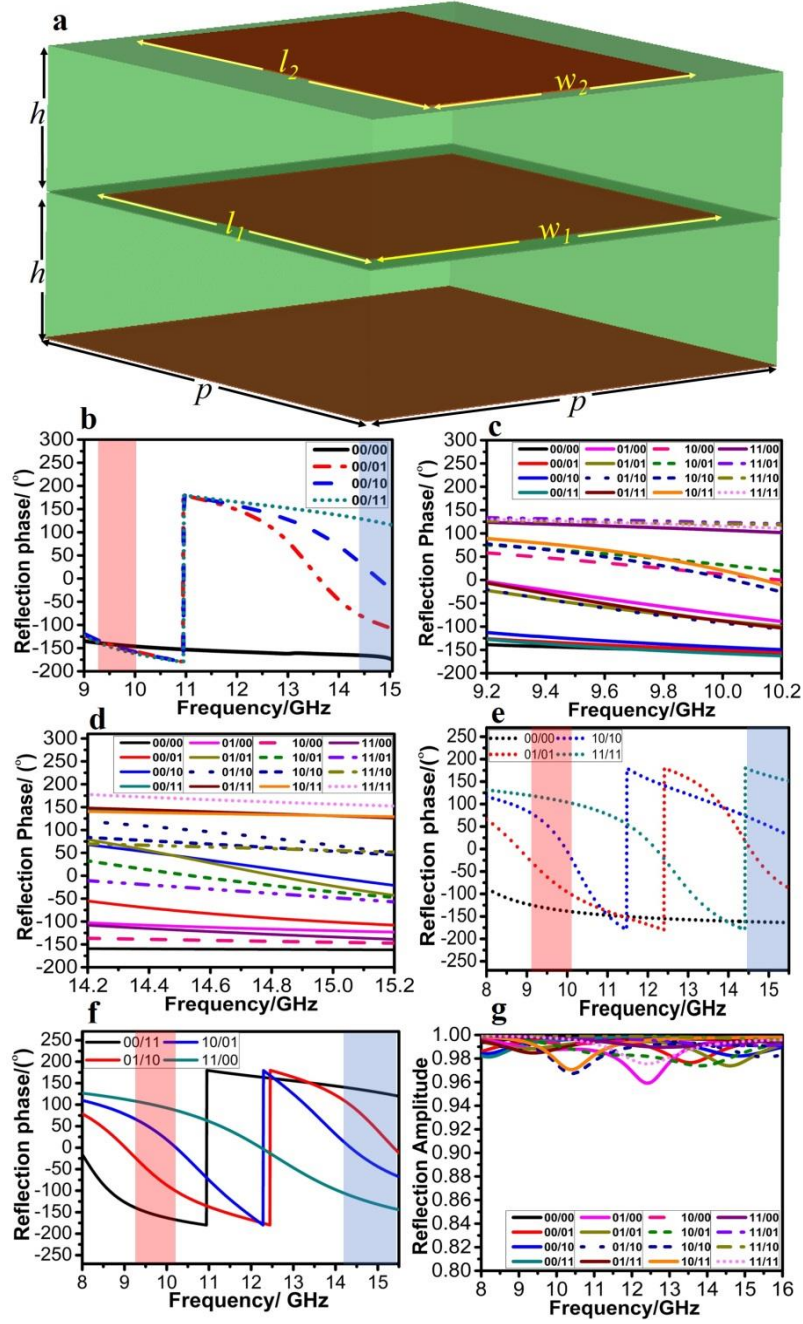

**Figure 2.** Demonstration of dual-band 2-bit unit-cell and their corresponding phase and amplitude responses. (**a**) Perspective transparent view of proposed unit-cell. (**b**) Phase response of first four meta-atoms in which the pink coloured shaded region represents the lower-band and blue-color is for higher-band. (**c**) Phase responses of all coding particles in lower-band. (**d**) Phase response of all coding particles in higher band. (**e**) Phase responses of diagonal particles of matrix '$M^{2-bit}$' for both operating bands. (**f**) Phase responses of off-diagonal particles of matrix '$M^{2-bit}$' for both operating bands. (**g**) Amplitude response of all coding particles in both operating bands.

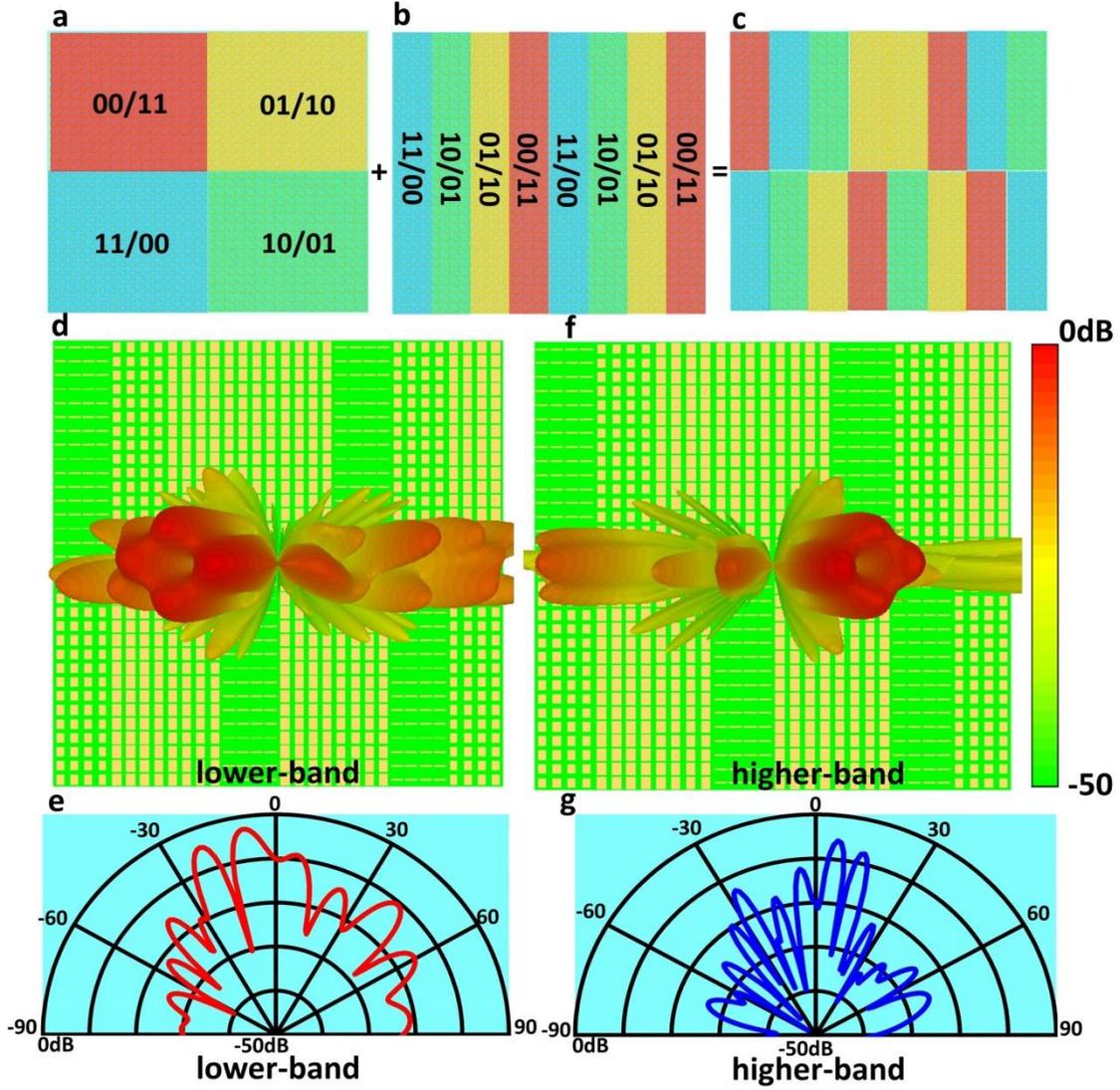

**Figure 3:** Coding patterns and the simulated 3D and 2D far-field scattering patterns to show the performance of dual-band OAM-VB steering for linear polarization incident with topological charge $l=1$. **(a)** Coding pattern of the normal OAM-VB with coding sequence '$M_1$'. **(b)** Coding pattern of the phase gradient with coding sequence '$M_2$'. **(c)** The mixed coding pattern '$M_3$' is formed by adding the coding patterns of (a) and (b). **(d-e)** 3D and 2D scattering pattern of OAM-VB steering in the lower-band for x-polarized incident at frequency of 10 GHz. In which the normalized incident wave creates OAM-VB that steers to left direction. **(f-g)** 3D and 2D scattering pattern of OAM-VB steering in the higher-band for x-polarized incident at frequency of 15 GHz. In which the normalized incident causes OAM-VB generation that steers to right direction.

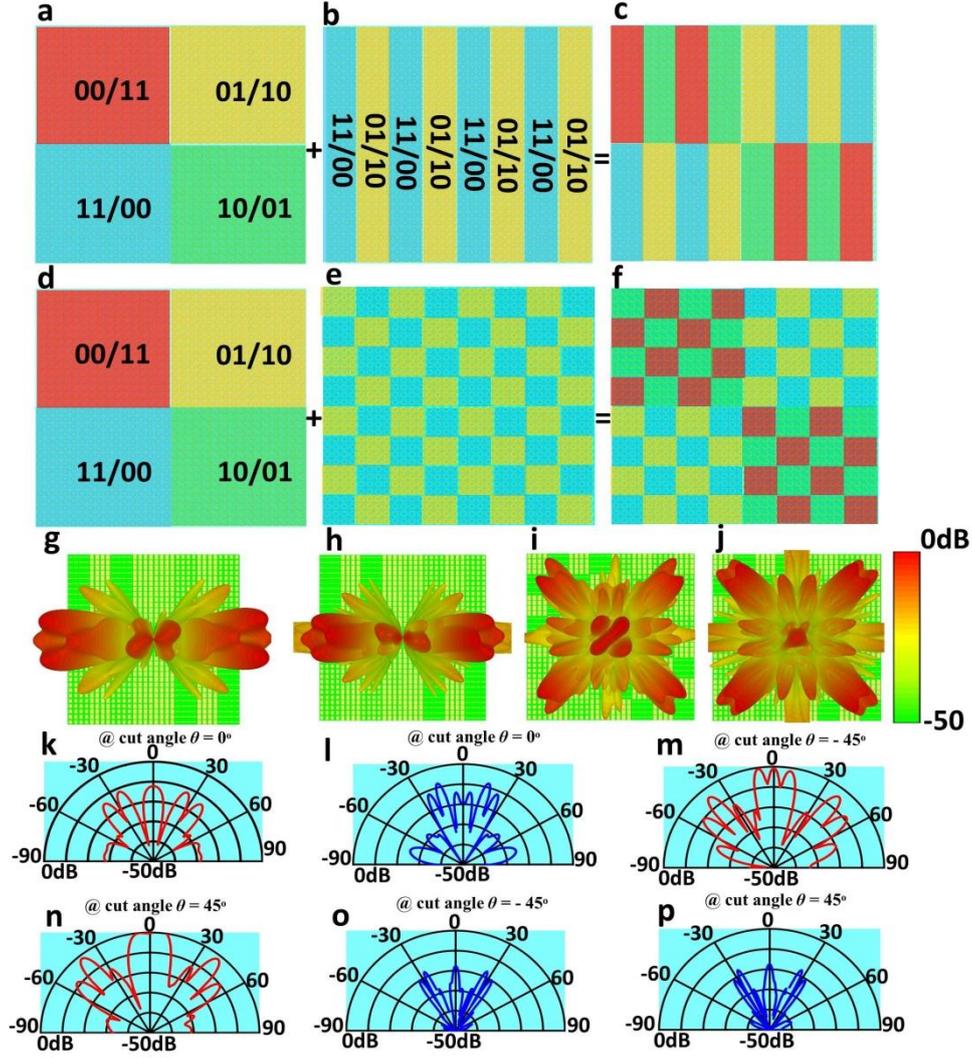

**Figure 4:** Coding patterns and the simulated 3D and 2D far-field scattering patterns to show the performance of dual-band OAM-VB splitting and OAM-VB quad-beam splitting for linear polarization incident with topological charge $l = 1$. **(a)** Coding pattern of the normal OAM-VB with coding sequence 01230123…. **(b)** Coding pattern of normal beam-splitting with periodic coding sequence 31313131…. **(c)** The mixed coding pattern formed by adding the coding patterns in (a) and (b). **(d-f)** Coding pattern of (a) is mixed with the coding sequence of quad-beam splitting (e) to form the proposed OAM-VB quad-beam splitting CM (f). **(g,k)** 3D and 2D scattering pattern of OAM-VB splitting in the lower-band for x-polarized incident. In which the normalized incident wave creates two different OAM-VB in opposite half-planes. **(h,l)** 3D and 2D scattering pattern of OAM-VB splitting in the higher-band for x-polarized incident. In which the normalized incident wave creates two different OAM-VB in opposite half-planes. **(i,m,n)** 3D scattering pattern of OAM-VB quad-splitting in the lower-band for x-polarized incident (i). In which the normalized incident wave creates four different OAM-VB in opposite half-planes. 2D scattering pattern of OAM-VB quad-splitting when $\theta = -45°$ (m) and $\theta = 45°$ (n). **(j,o,p)** 3D scattering pattern of OAM-VB quad-splitting in the higher-band for x-polarized incident (j). In which the normalized incident wave creates four different OAM-VB in opposite half-planes. 2D scattering pattern of OAM-VB quad-splitting when $\theta = -45°$ (o) and $\theta = 45°$ (p)

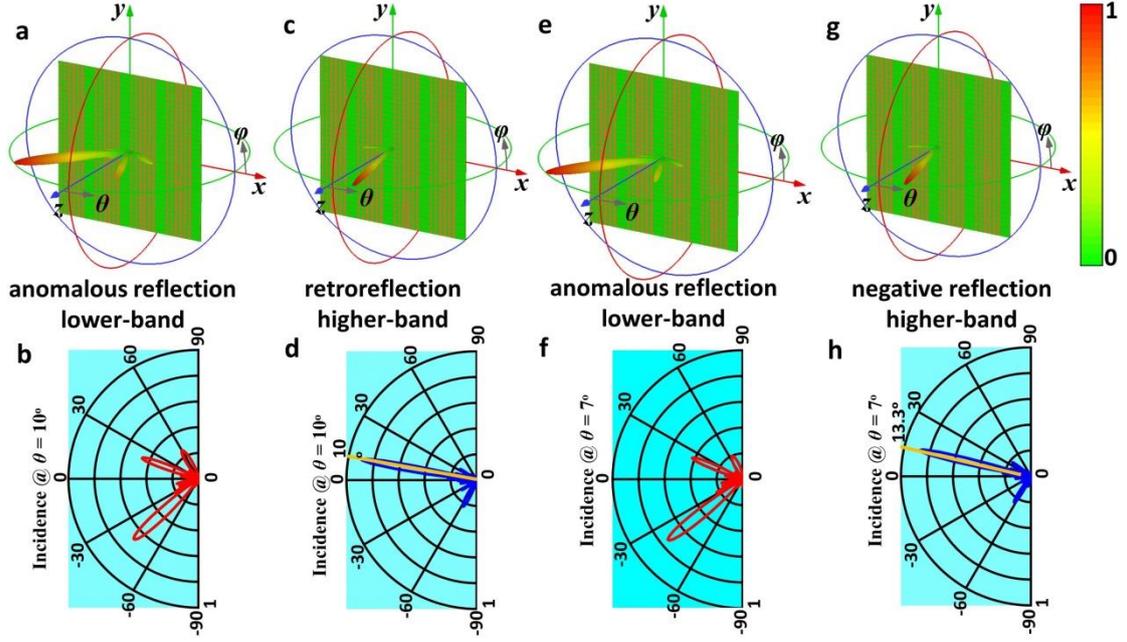

**Figure 5:** Simulated 3D and 2D far-field scattering patterns of proposed CM to generate anomalous reflection in lower-band along with the retroreflection and negative reflection at higher-band with oblique incidence. **(a,b)** The 3D and 2D scattering patterns of anomalous reflection in the lower-band for x-polarized oblique incidence at $\theta = 10°$, respectively. **(c,d)** The 3D and 2D scattering patterns of retroreflection in the higher-band for x-polarized oblique incidence at $\theta = 10°$, respectively. **(e,f)** The 3D and 2D scattering patterns of anomalous reflection in the lower-band for x-polarized oblique incidence at $\theta = 7°$, respectively. **(g,h)** The 3D and 2D scattering patterns of negative reflection in the higher-band for x-polarized oblique incidence at $\theta = 7°$, respectively.